\documentclass[12pt, a4paper]{article}
\usepackage{jheparxiv}
\usepackage{wasysym}
\usepackage[utf8]{inputenc}
\usepackage{amsmath}
\usepackage{amsfonts}
\usepackage{amssymb}
\usepackage{latexsym}
\usepackage{mathrsfs}
\usepackage{mathtools}
\usepackage{braket}		
\usepackage{graphicx}
\usepackage{color}
\usepackage{xcolor}
\usepackage{slashed}
\usepackage{twistor}
\usepackage[all]{xy}
\usepackage{tikz-cd}
\usepackage{bbm}

\newcommand{\beq}{\begin{equation}}
\newcommand{\eeq}{\end{equation}}
\newcommand{\bes}{\begin{subequations}}
\newcommand{\ees}{\end{subequations}}
\newcommand{\bee}{ \begin{align}}

\renewcommand{\d}{\mathrm{d}}

\newcommand{\eps}{\epsilon}
\newcommand{\Dbar}{\bar{D}}
\newcommand{\g}{\mathfrak{g}}

\newcommand{\vphi}{\varphi}
 
\renewcommand{\bz}{{\bar z}}

\newcommand{\Tr}{{\mathrm{Tr}}}

\newcommand{\bJ}{{\mathbf{J}}}

\newcommand{\Dhat}{{\hat{D}}}
\newcommand{\bm}{{\bar m}}

\title{Celestial 1-form symmetries}

\author[a]{Laurent Freidel,}
\author[b]{Atul Sharma}

\affiliation[a]{Perimeter Institute for Theoretical Physics,\\
31 Caroline Street North, Waterloo, ON, N2L 2Y5, Canada}
\emailAdd{lfreidel@pitp.ca}

\affiliation[b]{Center for the Fundamental Laws of Nature \& Black Hole Initiative,\\Harvard University, Cambridge, MA, 02138, USA \vspace{0.1cm}}
\emailAdd{atulsharma@fas.harvard.edu}

\abstract{The $S$-algebra originally arose as a chiral algebra of asymptotic symmetries of Yang-Mills theory. We show that in the self-dual sector of Yang-Mills, the $S$-algebra gets upgraded to an infinite-dimensional algebra of $1$-form symmetries in the bulk. The associated 2-form currents encode the integrability and hierarchies of self-dual Yang-Mills. As an application, we prove the equality of Carrollian corner charges with modes of the celestial chiral algebra by expressing them as integrals of the same 2-form currents over homologous 2-cycles.}

\begin{document}

\maketitle

\section{Introduction}
\label{sec:intro}

Symmetries play a pivotal role in the study of quantum field theories. Recent years have witnessed the discovery of numerous \emph{generalized symmetries}, which provide a powerful new tool to capture their non-perturbative aspects \cite{Gaiotto:2014kfa, Gaiotto:2017zba,Bhardwaj:2023kri,Iqbal:2024pee}. Higher-form symmetries are now regularly utilized to constrain the spectrum of extended operators and defects. Their breaking is also useful for classifying vacua and detecting phase transitions such as the deconfinement transition in gauge theory, see \emph{e.g.} \cite{Gaiotto:2017yup,Komargodski:2020mxz}.

A parallel line of research has focused on the study of \emph{asymptotic symmetries}. Bottom-up studies of holographic dualities have resulted in the discovery of infinite-dimensional asymptotic symmetry algebras in a variety of settings. They underlie the soft theorems that capture the universal infrared physics of scattering amplitudes in asymptotically flat spacetimes \cite{Strominger:2017zoo}. They also provide a bottom-up tool to count the microstates of asymptotically AdS black holes in low dimensions \cite{Brown:1986nw,Strominger:1997eq}.

Both types of symmetries share one fundamental feature in common: their generators are obtained as integrals over a codimension-$2$ surface, rather than as integrals of current densities over a codimension-$1$ slice.
The difference is that for an asymptotic symmetry, the Noether charge is glued to the asymptotic boundary and is usually interpreted as a $0$-form symmetry for the boundary theory. Therefore, beyond some very recent investigations \cite{Berean-Dutcher:2025ohp,Tizzano:2026rgr}, these two formalisms have had a largely disconnected history.

Nonetheless, it is natural to look for settings in which they overlap in interesting ways. In this work, we show that self-dual gauge theory provides such a setting. 
One can free the asymptotic charge from its gluing to the asymptotic boundary and promote it to a bona fide $1$-form symmetry! 

One surprising outcome is that the 1-form symmetries revealed in self-dual Yang-Mills form a noncommutative algebra, \emph{viz.}, the celebrated $S$-algebra \cite{Guevara:2021abz,Strominger:2021mtt}. 
This result generalizes the standard expectation that $1$-form symmetry generators should be commutative \cite{Gaiotto:2014kfa, Gaiotto:2017zba,  Bhardwaj:2023kri,  Iqbal:2024pee}. As we will see, this follows from the understanding of asymptotic charges as corner charges \cite{Donnelly:2016auv, Ciambelli:2022vot} that survive the imposition of the gauge constraint in the bulk of the slice. Such charges associated with a given codimension-$2$ surface are therefore implicitly dependent on which codimension-$1$ slice they are connected to. 

The language of 1-form symmetries also provides a unifying link between the celestial \cite{Pasterski:2021raf} and Carrollian \cite{Ruzziconi:2026bix} approaches to flat space holography. Working in 4D flat space, the celestial approach reinterprets asymptotic symmetries as modes of conserved currents of a 2D CFT living on the celestial sphere. These modes are given by integrals over Bondi time $u$ and contour integrals on the celestial sphere, see \emph{e.g.} \cite{Pasterski:2022jzc,Pano:2023slc,Strominger:2017zoo} and references therein. On the other hand, the Carrollian approach views them as symmetries of a 3D theory living on null infinity, whose charges are integrals over the whole celestial sphere at fixed time $u$, see e.g. \cite{Strominger:2013jfa, Freidel:2021ytz, Cresto:2025ubl}. 

The twistorial approach of \cite{Kmec:2025ftx} showed that both types of charges were built out of the same building blocks, the charge aspects of \cite{Freidel:2023gue}. Our 1-form symmetry analysis provides a purely spacetime proof of the fact that the celestial and corner charges are actually \emph{equal} to each other. It accomplishes this by showing that both types of charges are integrals of the same conserved 2-forms over different choices of 2-cycles, and that the chosen 2-cycles are homologous to each other. This provides further evidence for the equivalence of the celestial and Carrollian approaches anticipated in \cite{Donnay:2022aba,Donnay:2022wvx}.

We begin in section \ref{sec:sdym} with a brief review of self-dual Yang-Mills (sdYM) theory and its Lax formulation. This is the theory of a nonlinear, self-dual (sd) gauge field $A$, and a 2-form $B$ describing an anti-self-dual (asd) linear perturbation. In section \ref{sec:R}, we review the recursion operator of sdYM, which is a hallmark of its integrability. It recursively generates a series of on-shell asd 2-forms $B_s$ from a seed asd field $B_0 = B$. In section \ref{sec:2form}, we construct 2-form symmetry currents of sdYM of the form
\be
\bJ_\tau = \sum_{s\in\Z}\Tr(\tau_sB_s)
\ee
labeled by symmetry parameters given by scalar functions $\tau_s$. We prove that $\d\bJ_\tau=0$ when the $\tau_s$ are related to each other by the inverse of the same recursion operator. This furnishes an infinite class of 1-form symmetries. 

Section \ref{sec:max} briefly discusses the sd Maxwell case, relating soft photon asymptotic symmetries to the 1-form symmetries of electromagnetism. In section \ref{sec:S}, we generalize to the non-abelian case by relating our 1-form symmetries to the $S$-algebra. Conserved charges of the $S$-algebra are generated by integrating our 2-form currents on codimension-2 cycles contained within null infinity. Here, we also prove the equivalence between the celestial and Carrollian notions of asymptotic symmetry charges. Finally, in section \ref{sec:discussion}, we end with a brief discussion of the noncommutativity of our 1-form symmetries.


\section{Self-dual gauge theory}
\label{sec:sdym}

 Self-dual Yang-Mills (sdYM) is an integrable subsector of Yang-Mills theory \cite{Mason:1991rf}. Consider a connection $A\in\Omega^1(M,\mathfrak{g})$ on a 4-manifold $(M,g)$, valued in a Lie algebra $\mathfrak{g}$. Let $\D = \d + A$ be the gauge covariant derivative, with curvature $F=\d A+\frac12\,[A,A]$. The action for sdYM is given by
\be
S_\text{sdYM} = \int_M \Tr\,B\wedge F_-\,,
\ee
where $F_-$ denotes the anti-self-dual (asd) part of $F$. The field $B\in\Omega^2_-(M,\mathfrak{g})$ is an asd Lagrange multiplier whose equation of motion imposes self-duality: $F_- = 0$. The equation of motion of $A$ takes the form of a Bianchi identity: $ \D B = 0$.

We will study this theory on Minkowski space in planar Bondi gauge,\footnote{The gauge field $A$ must be complexified when working in Lorentzian signature, but this does not pose a problem in perturbation theory.}
\be\label{flat}
\d s^2 = -2\,\d u\,\d r + 2\,r^2\,\d z\,\d\bz\,,
\ee
where $u,r\in\R$ are real and $z\in\C$ is a complex coordinate. In these coordinates, the future and past null infinities $\scri^\pm$ are approached as $r\to\pm\infty$. The action of sdYM is conformally invariant, so we introduce a new coordinate $v = r^{-1}$ and work with the conformally equivalent metric
\be
\d\tilde s^2 = v^{2}\d s^2 = 2\,\d u\,\d v + 2\,\d z\,\d\bz\,,
\ee
which is the product metric of $\R^{1,1}\times\C$. The boundaries $\scri^\pm$ are now approached as one sends $v\to0^\pm$. In what follows, we will collectively denote the coordinates $(u,v,z,\bz)$ by $x^\mu$. 

In the orientation picked by the volume form $\d u\wedge\d v\wedge\d z\wedge\d\bz$, the three independent components of the self-duality equation $F_-=0$ are given by
\be\label{sdeq}
F_{v\bz} = F_{uz} = F_{uv}+F_{z\bz}=0\,.
\ee
The integrability of sdYM is encapsulated in the existence of the Lax pair,
\be\label{Lax}
\begin{split}
    &L = qD_u + D_\bz\,,\\
    &M = qD_z - D_v\,,
\end{split}
\ee
where $D_\mu=\p_\mu+A_\mu$, and $q\in\CP^1$ is the Lax parameter. The commutator of $L$ and $M$ is given by
\be
[L,M] = F_{v\bz}-q(F_{uv}+F_{z\bz})+q^2F_{uz}\,,
\ee
which vanishes for all $q$ if and only if \eqref{sdeq} holds.

The equations of self-duality also take a beautiful guise when written in terms of (pseudo-)complex geometry. To describe this, we introduce a Lorentzian analog of the quaternionic Dolbeault complex of \cite{verbitsky2006} by defining the exterior derivatives
\be
\begin{split}
    \dbar &= \d v\,\p_v + \d\bz\,\p_\bz\,,\\
    \dhat &= \d v\,\p_z - \d\bz\,\p_u\,.
\end{split}
\ee
These satisfy $\dbar^2=\dhat^2=[\dbar,\dhat]=0$. We can covariantize these by replacing ordinary derivatives by covariant derivatives,
\be\label{D+-}
\begin{split}
    \bar D &= \d v\,D_v + \d\bz\,D_\bz\,,\\
    \hat D &= \d v\,D_z - \d\bz\,D_u\,.
\end{split}
\ee
Then \eqref{sdeq} are equivalent to the zero curvature conditions
\be
\Dbar^2 = 0\,,\qquad \Dhat^2 = 0\,,\qquad [\Dbar,\Dhat] = 0\,.
\ee
The operators $\Dbar,\Dhat$ will arise naturally in the coming sections.


\section{Charge aspects}
\label{sec:R}

In our choice of orientation, the space of asd 2-forms is spanned by the triplet
\be
    \Sigma_- = \d u\wedge\d z\,,\qquad\Sigma_0 = \d u\wedge\d v+\d z\wedge\d\bz\,,\qquad\Sigma_+ = \d v\wedge\d\bz\,.
\ee
For instance, we can expand the asd field $B$ as follows:
\be
B = B_{uz}\,\Sigma_- + B_{z\bz}\,\Sigma_0 + B_{v\bz}\,\Sigma_+\,,
\ee
having employed the anti-self-duality constraints $B_{u\bz} = B_{vz} = B_{uv}-B_{z\bz}=0$. 

We now introduce a set of \emph{charge aspects} out of which we will build our conserved 2-form currents. To this end, let us rename the three independent components of the asd field $B$ as follows:
\be\label{RB}
R_{-1} = B_{uz}\,,\qquad R_0 = B_{z\bz}\,,\qquad R_1 = B_{v\bz}\,.
\ee
In this notation, the Gauss Law components $D_{[\mu}B_{\nu\rho]}=0$ can be written as
\be\label{DB}
\begin{split}
    &D_uR_s + D_\bz R_{s-1} = 0\,,\\
    &D_zR_s - D_vR_{s-1} = 0\,.
\end{split}
\ee
Here, the range of $s$ is $s=0,1$, but we can choose to continue this pattern.

To do this, note that because $L$ and $M$ commute on the support of $F_-=0$, we can look for a simultaneous solution $b(x,q)$ of the equations
\be\label{laxeq}
Lb = Mb = 0\,,
\ee
where $L,M$ act on $b$ in the adjoint representation. Let us solve this on the patch $\{q\neq0,\infty\}\subset\CP^1$. 

On this annulus, we expand $b(x,q)$ as a Laurent series in $q$,
\be\label{bR}
b(x,q) = \sum_{s\in\Z}\frac{R_s(x)}{q^{s+2}}\,.
\ee
The Lax equations \eqref{laxeq} translate into the following recursion relations on the coefficients $R_s$,
\be\label{rec}
\begin{split}
    &D_uR_s + D_\bz R_{s-1} = 0\,,\\
    &D_zR_s - D_vR_{s-1} = 0\,,
\end{split}
\ee
for all $s\in\Z$. For $s=0,1$, these reproduce the Bianchi identity $ \D B=0$. So we may initiate the recursion for either $s<0$ or $s>1$ by viewing \eqref{RB} as a boundary condition.
More compactly, we can express these recursion relations as
\be\label{rec1}
\Dhat R_s = \Dbar R_{s-1}\qquad\forall\;s\in\Z\,,
\ee
in terms of the differential operators mentioned in \eqref{D+-}. Because $\Dhat^2=0$, we find the consistency condition
\be
\Dhat\Dbar R_s = 0\implies (D_u D_v + D_z D_\bz)R_s = 0\,.
\ee
So every $R_s$ solves the gauge-covariant wave equation.

This is the spacetime reconstruction of a \v{C}ech representative $b$ that uplifts $B$ to twistor space $\PT$. Indeed, we can verify the Penrose contour integral formulae \cite{Penrose:1972ia},
\be\label{penrose}
B_{uz} = \oint_{q=0}\d q\,b\,,\qquad B_{z\bz} = \oint_{q=0}\d q\,q\,b\,,\qquad B_{v\bz} = \oint_{q=0}\d q\,q^2b\,.
\ee
More precisely, the trivialization of $b$ in a frame that is covariantly constant with respect to $L,M$ provides the representative of a class in $\check{H}^1(\PT,\CO(-4)\otimes\g)$. The Penrose transform guarantees the existence of a unique such representative $b$ for every on-shell asd field $B$ \cite{Eastwood:1981jy,Ward:1990vs}.

We will refer to the quantities $R_s$ as charge aspects. They act as building blocks for our higher-form conserved currents. Their existence is a consequence of the integrability of sdYM. As explained in \cite{Mason:1991rf,Kmec:2025ftx}, the operation $\mathscr{R}$ on the space of solutions of the covariant wave equation given (up to constants of integration) by
\be
R_s = \mathscr{R}R_{s-1}\,,\qquad \mathscr{R} \vcentcolon= \Dhat^{-1}\circ\Dbar\,,
\ee
defines the \emph{recursion operator} of sdYM. The resulting $R_s$ give rise to a hierarchy of commuting flows on the space of solutions of the self-duality equations. 

The aspects $R_s$ obtained by this procedure can be arranged into a series of new asd 2-forms that we will refer to as the ``radiation 2-forms'',
\be\label{Bs}
B_s = R_{s-1}\Sigma_- + R_s\Sigma_0 + R_{s+1}\Sigma_+\,.
\ee
Note in particular that $B_0 = B$, the original Lagrange multiplier 2-form. Due to \eqref{rec}, every $B_s$ solves the Gauss law $\D B_s = 0$, and its \v{C}ech twistor representative $b_s$ is obtained from that of $B_{s-1}$ by the uplift of the recursion operator to twistor space: $b_s = \mathscr{R}b_{s-1}\vcentcolon=q\,b_{s-1}$ \cite{Mason:1991rf,Dunajski:2000iq}. Solving this recursion gives $b_s = q^sb$. Replacing $b$ by this $b_s$ in the Penrose integrals \eqref{penrose} is readily seen to reproduce the components of $B_s$.


\section{1-form symmetries}
\label{sec:2form}

In this section, we will use judicious combinations of the charge aspects $R_s$ to generate conserved 2-form currents of sdYM. Our construction is analogous to the construction of conserved currents associated with the isometries of a metric. In QFT on curved spacetime, if $\xi^\mu$ is a vector field on a spacetime $(M,g)$, and $T_{\mu\nu}$ is the stress tensor, then the corresponding 1-form $j_\mu = T_{\mu\nu}\xi^\nu$ is a conserved current precisely when $\xi^\mu$ is Killing. In our case, the charge aspects $R_s$ are analogous to the stress tensor, and we will look for symmetry parameters $\tau_s$ that can pair with the charge aspects to generate 2-form conserved currents.

Pick a set of $\g$-valued functions $\tau_s(x)$, $s\in\Z$, to play the role of our symmetry parameters. As we did with $b(x,q)$ in \eqref{bR}, arrange the $\tau_s$ in a Laurent series in $q$,
\be\label{tauq}
\tau(x,q) = \sum_{s\in\Z}\tau_s(x)\,q^s\,.
\ee
We propose to pair $\tau(x,q)$ with $b(x,q)$ to build a 2-form current as a contour integral in $q$ around $q=0$,
\be\label{Jcont}
\bJ_\tau = \oint_{q=0}\Tr(\tau b)\;\d q\wedge\Sigma\,.
\ee
Here, $\Sigma$ is the Gindikin 2-form,
\be
\Sigma = \Sigma_- + q\Sigma_0 + q^2\Sigma_+\,.
\ee
This acts as an $\CO(2)$-valued symplectic structure on the fibers of the twistor fibration $\PT = \mathrm{Tot}(\CO(1)\oplus\CO(1)\to\CP^1)$, and $\d q\wedge\Sigma$ acts as an $\CO(4)$-valued holomorphic top-form on $\PT$.

Evaluating the contour integral yields the asd 2-form
\be
\bJ_\tau = J_\tau^-\Sigma_- + J_\tau^0\Sigma_0 + J_\tau^+\Sigma_+\,,
\ee
with coefficients that take the form
\be
    J_\tau^\pm = \sum_{s\in\Z}\Tr(\tau_sR_{s\pm1})\,,\qquad J_\tau^0 = \sum_{s\in\Z}\Tr(\tau_sR_{s})\,.
\ee
We can also write $\bJ_\tau$ as
\be\label{Jt}
\bJ_\tau = \sum_{s\in\Z}\Tr(\tau_s B_s)\,,
\ee
in terms of the radiation 2-forms obtained in \eqref{Bs}. A crucial point is that one can often make choices for $\tau_s$ so that these expressions reduce to finite sums or sums over $s\geq0$. 

\medskip

We now try to find the conditions on $\tau$ under which the 2-form $\bJ_\tau$ is closed. In brief, we will find that $\tau$ will also have to satisfy the Lax equations $L\tau=M\tau=0$.

A short computation tells us that
\be\label{dJtau}
    \d\bJ_\tau = (\dbar J_\tau^- - \dhat J_\tau^0)\wedge\Sigma_- + (\dbar J_\tau^0 - \dhat J_\tau^+)\wedge\Sigma_0\,.
\ee
 So we need to find the conditions under which the coefficients of $\Sigma_-,\Sigma_0$ in this expression vanish. It is worth computing one of these in detail. For instance, we find that
 \begin{align}
     \dbar J_\tau^- - \dhat J_\tau^0 &= \sum_{s\in\Z}\Tr\big(\Dbar\tau_s\,R_{s-1} + \tau_s\,\Dbar R_{s-1} - \Dhat\tau_s\,R_s - \tau_s\,\Dhat R_s\big)\nonumber\\
     &= \sum_{s\in\Z}\Tr\big[\big(\Dbar\tau_s - \Dhat\tau_{s-1}\big)R_{s-1}\big]\,,\label{dJ-}
 \end{align}
 having used the recursion $\Dhat R_s = \Dbar R_{s-1}$ to get the second line. An identical calculation yields
 \be\label{dJ0}
\dbar J_\tau^0 - \dhat J_\tau^+ = \sum_{s\in\Z}\Tr\big[\big(\Dbar\tau_s - \Dhat\tau_{s-1}\big)R_s\big]\,.
\ee
 Collecting together \eqref{dJtau}--\eqref{dJ0}, we obtain the evolution of $\bJ_\tau$,
\be\label{dJ}
\d\bJ_\tau = \sum_{s\in\Z}\Tr\big[\big(\Dbar\tau_s - \Dhat\tau_{s-1}\big)\wedge B_s\big]\,,
\ee
written in terms of the radiation 2-forms given in \eqref{Bs}.

From \eqref{dJ}, we conclude that we obtain
\be
\d\bJ_\tau=0
\ee
if and only if our symmetry parameters $\tau_s(x)$ obey the \emph{dual} recursion relations,
\be
\Dbar\tau_s = \Dhat\tau_{s-1}\qquad\forall\;s\in\Z\,.
\ee
In components, these read
\be\label{dualrec}
\begin{split}
    &D_\bz\tau_s + D_u\tau_{s-1} = 0\,,\\
    &D_v\tau_s - D_z\tau_{s-1} = 0\,.
\end{split}
\ee
It is easily checked that these are equivalent to demanding that $\tau(x,q)$ obeys the Lax equations $L\tau=M\tau=0$ for all $q\neq0,\infty$. More formally, $\tau(x,q)$ can be viewed as a section of $\check{H}^0(\PT',\g)$, where $\PT'=\PT-\{q=0,\infty\}$. In twistor theory, this arises as a large gauge transformation with singularities at $q=0,\infty$ \cite{Kmec:2025ftx}.

Every set of symmetry parameters $\tau_s$ obeying these equations defines a conserved 2-form on spacetime. These 2-forms give rise to 1-form symmetries of sdYM on flat space. Given a choice of codimension-2 closed surface $C$, the charge associated to $\bJ_\tau$ is given by
\be
Q_\tau(C) = \int_C\bJ_\tau\,.
\ee
The closure of $\bJ_\tau$ ensures that this expression is conserved under deformations of $C$. 


\section{Example: sd Maxwell} 
\label{sec:max}

The simplest example of this construction arises in self-dual Maxwell (sdM) theory. In that case, the adjoint representation is trivial. So the Lax equations on $\tau(x,q)=\sum_s\tau_sq^s$ reduce to
\be
(q\p_u+\p_\bz)\tau = (q\p_z-\p_v)\tau = 0\,.
\ee
These can be solved by the method of characteristics to obtain
\be
\tau = f(q,u-q\bz,z+qv)
\ee
for some arbitrary function $f$ of three complex variables. As a simple situation, we will assume that $f(q,u-q\bz,z+qv)$ admits a Taylor expansion around $q=0$. Then we can take the $\tau_s$ for $s\geq0$ to be the coefficients in this Taylor expansion, and $\tau_s = 0$ for $s<0$.

At the same time, up to constants of integration, we can solve the recursion for $R_s$ for $s>1$ starting from an initial configuration of the abelian asd field $B$. Beyond the initialization \eqref{RB}, in the abelian case, the first recursion in \eqref{rec} yields
\be
R_{s+1} = (-\p_u)^{-s}\p_\bz^sB_{v\bz}\,,\qquad s\geq1\,.
\ee
Consistency with the second recursion in \eqref{rec} holds because $B_{v\bz}$ obeys the wave equation $(\p_u\p_v+\p_z\p_\bz)B_{v\bz}=0$. This procedure generates a series of asd 2-forms $B_s$, $s\geq0$, obeying $\d B_s=0$,
\be
\begin{split}
    &B_0 = B = B_{uz}\,\Sigma_- + B_{z\bz}\,\Sigma_0 + B_{v\bz}\,\Sigma_+\,,\\
    &B_1 = B_{z\bz}\,\Sigma_- + B_{v\bz}\,\Sigma_0 - \p_u^{-1}\p_\bz B_{v\bz}\,\Sigma_+\,,\\
    &B_{s+1} = [(-\p_u)^{1-s}\p_\bz^{s-1}B_{v\bz}]\,\Sigma_- + [(-\p_u)^{-s}\p_\bz^{s}B_{v\bz}]\,\Sigma_0+ [(-\p_u)^{-s-1}\p_\bz^{s+1}B_{v\bz}]\,\Sigma_+
\end{split}
\ee
where the last line is valid for $s\geq 1$.
In these expressions, the $u$ integrals are performed along null rays of fixed $v,z,\bz$.

Pairing $\tau_s$ with $R_s$, we obtain the conserved 2-forms
\be
\bJ_\tau^\text{sdM} = \sum_{s=0}^\infty\tau_sB_s\,.
\ee
The $\tau_s = \delta_{s,0}$ case is the simplest, for which we get
\be
\bJ^\text{sdM}_{\tau=1} = B\,.
\ee
This recovers an asd combination of the electric and magnetic 1-form symmetries of Maxwell theory. It is also possible to recover the sd combination using the negative helicity charge aspects from \cite{Kmec:2025ftx}, to which we hope to return in the future. 

More novel are the cases of non-constant $\tau$. For example, we could take $\tau$ to be a function purely of the single variable $z+qv$, say
\be
\tau = \eps(z+qv)\,.
\ee
Expanding this around $q=0$ to extract $\tau_s$, we obtain the conserved 2-forms
\be\label{jmax}
\bJ^\text{sdM}_{\tau=\eps} = \sum_{s=0}^\infty\frac{v^s}{s!}\,\p_z^s\eps(z)B_s\,.
\ee
As we have seen, these currents owe their existence to the integrability and hierarchies of the self-dual theory which underlie the construction of $B_s$. 

At this stage, we are ready to relate these to the leading soft photon asymptotic symmetries of sd Maxwell theory \cite{He:2014cra}. If we restrict \eqref{jmax} to, say, future null infinity $v\to0^+$, we obtain
\be
\lim_{v\to0^+}\bJ^\text{sdM}_{\tau=\eps} = \eps(z)B\bigr|_{\scri^+}\,.
\ee
Integrating this over a light cone cut $S^2_u$ of fixed $u$ on $\scri^+$, we get the conserved charges
\be
Q_\eps = \int_{S^2_u}\eps(z)B\,.
\ee
This makes contact with the recent investigations \cite{Berean-Dutcher:2025ohp,Tizzano:2026rgr} of the connections between asymptotic and 1-form symmetries in electromagnetism. More general $\tau$ will produce soft symmetries associated to subleading soft photons. We will study this as well as non-abelian generalizations in the next section.


\section{The $S$-algebra}
\label{sec:S}

To find non-abelian examples of the general construction described in section \ref{sec:2form}, we study the case of the $S$-algebra familiar from celestial holography. This algebra originally arose in the collinear limits of gluon scattering amplitudes \cite{Guevara:2021abz,Strominger:2021mtt}. It was interpreted as an asymptotic symmetry algebra of gauge theory in flat space, and its Noether charges were obtained in \cite{Freidel:2023gue,Cresto:2025bfo}. 

These charges arose as ordinary 0-form generators of large gauge symmetries: integrals over 3D Cauchy surfaces that reduced to boundary integrals over 2-spheres at null infinity. We now use the construction from the previous section to argue that in \emph{self-dual} gauge theory, it is possible to replace these boundary 2-spheres by any 2-sphere in the bulk.

Our first task is to find a basis of solutions of the dual recursion relations \eqref{dualrec} that give rise to the $S$-algebra symmetries. This proceeds as follows. Let us work in the patch $q\neq\infty$. In this patch, we can find a frame $U(x,q)$ that is covariantly constant with respect to $L,M$,
\be
LU = MU = 0\,,
\ee
where $L,M$ act in the fundamental representation from the left. As previously mentioned, the dual recursion is equivalent to the Lax equations $L\tau=M\tau=0$. Dressing $\tau$ with the frame $U$, a standard calculation shows that these are solved by
\be
\tau = U f U^{-1}\,,\qquad f\equiv f(q,u-q\bz,z+qv)\,.
\ee
So we reduce the problem to prescribing free functions $f$ of three variables.

Motivated by past explorations such as \cite{Kmec:2025ftx}, we choose to work with the following class of functions,
\be
f^{p,a}_{m,\bm} = \frac{(u-q\bz)^{p+\bm-1}q^{p-\bm-1}}{(z+qv)^m}\,\mathfrak{t}^a\,,
\ee
where $p=\frac32,1,\frac52,\dots$;   $1-p\leq \bm\leq p-1$ with $m\in\Z$, and $\mathfrak{t}^a$ are Lie algebra generators. Here, we have additionally restricted ourselves to the patch $q\neq-z/v$. The cases studied in \cite{Kmec:2025ftx} were restricted to null infinity $v=0$, for which these symmetry parameters simplified to
\be
f^{p,a}_{m,\bm}\bigr|_{v=0} = \frac{(u-q\bz)^{p+\bm-1}q^{p-\bm-1}}{z^m}\,\mathfrak{t}^a\,.
\ee
These were then interpreted as modes of a current in a celestial CFT on the $z$-plane. Our construction generalizes them to nonzero values of $v$ that can be arbitrarily deep inside the bulk.

As we are working in the patch $q\neq\infty$, we may expand the representative $f^{p,a}_{m,\bm}$, the frame $U$, and its inverse $U^{-1}$ as Taylor series in $q$ around $q=0$. So the modes of $\tau = U f^{p,a}_{m,\bm}U^{-1}$ in its $q$-expansion can be taken to be non-negative. They are picked out by associated contour integrals,
\be
\tau^{p,a}_{m,\bm,s} = \oint_{q=0}\frac{\d q}{q^{s+1}}\,U f^{p,a}_{m,\bm}U^{-1}\,.
\ee
Plugging these into our conserved 2-form currents \eqref{Jt} and integrating over any choice of a 2-cycle $C$ embedded in our spacetime, we obtain bulk extensions of the $S$-algebra charges,
\be
S^{p,a}_{m,\bm} = \int_{C}\sum_{s=0}^\infty\Tr(\tau^{p,a}_{m,\bm,s}B_s)\,.
\ee
Making them more explicit requires computing $U$. This can always be done perturbatively for ``small'' gauge fields $A$, resulting in Wilson-line type expressions \cite{Ward:1990vs}.

Without loss of generality, we can take $C$ to be the 2-sphere at fixed $u,v$, denoted $S^2_{u,v}$. The integrals as written above are a bit formal. The small-$q$ expansion of $f^{p,a}_{m,\bm}$ can generate singularities at $z=0,\infty$. So our integrals are potentially divergent near these points. A $z=0$ singularity can be regulated by replacing
\be
\frac{1}{z}\mapsto\frac{\bz}{|z|^2+\eps^2}\,,\qquad\eps\to0\,.
\ee
A similar trick works for singularities at $z=\infty$.

As we send $v\to0^+$, the sphere $S^2_{u,v}$ continuously deforms into the sphere $S^2_u$ describing the light cone cut at a given $u$ on $\scri^+$. As these two spheres are homologous, we find the same charges $S^{p,a}_{m,\bm}$ when using either of them as our 2-cycle (in the absence of operators obstructing the deformation of $S^2_{u,v}$ into $S^2_u$). Working with $S^2_u$ recovers the standard description of \emph{corner charges} for the $S$-algebra \cite{Freidel:2023gue,Cresto:2025bfo},
\be\label{corner}
S^{p,a}_{m,\bm} = \lim_{v\to0^+}\int_{S^2_{u,v}}\d^2z\sum_{s=0}^\infty\Tr(\tau^{p,a}_{m,\bm,s}R_s)\,.
\ee
The resulting generators obey the $S$-algebra brackets,
\be\label{Salg}
[S^{p,a}_{m,\bm},S^{q,b}_{n,\bar n}] = f^{ab}{}_c\,S^{p+q-1,c}_{m+n,\bm+\bar n}\,,
\ee
where $f^{ab}{}_c$ are the structure constants of $\g$.

Alternatively, we could choose $C$ to be a surface of constant $v$ and $|z|=1$. Such surfaces arise as boundaries of surfaces of constant Rindler time \cite{Pasterski:2022jzc,Crawley:2021ivb}. As we send $v\to0^+$, we obtain an integral over $u$ as well as over $z\in S^2_u$ along a contour surrounding, say, $z=0$. This yields the \emph{celestial charges},
\be\label{celestial}
S^{p,a}_{m,\bm} = \lim_{v\to0^+}\int_{-\infty}^\infty\d u\oint_{z=0}\d z\sum_{s=0}^\infty\Tr(\tau^{p,a}_{m,\bm,s}R_{s-1})\,.
\ee
Here, the independence of the $z$-integral from the choice of contour is also a consequence of the conservation of the 2-form integrand. Equivalently, as seen in \cite{Kmec:2025ftx}, one can also directly verify that the integrand is holomorphic in $z$ as long as the aspects die off at $u=\pm\infty$.

Since we are dealing with conserved currents, the resulting celestial charges coincide with the corner charges \eqref{corner}. This is because after removing the point $z=0$, we can continuously deform the Carrollian 2-cycle $\{u=v=0\}$ into the celestial 2-cycle $\{v=0,|z|=1\}$: they form the boundaries at $\al=0,1$ of the 3-manifold,
\be
v = 0\,,\qquad u = \al t\,,\qquad z = ((1-\al)\e^t+\al)\,\e^{i\vphi}\,,
\ee
where the parameters $\al,t,\vphi$ range over $\al\in[0,1]$, $t\in\R$, $\vphi\in S^1$. 

As a result, the two types of charges will obey the same algebra \eqref{Salg}. From these celestial charges, one can construct celestial CFT currents
\be
S^{p,a}_\bm(z) = \sum_{m\in\Z}\frac{S^{p,a}_{m,\bm}}{z^{m+1}}
\ee
which are found to obey the operator product algebra
\be
S^{p,a}_\bm(z)\,S^{q,b}_{\bar n}(w) \sim \frac{f^{ab}{}_c\,S^{p+q-1,c}_{\bm+\bar n}(w)}{z-w}\,.
\ee
This is the OPE algebra of positive helicity soft gluons.

This establishes another equivalence between the Carrollian and celestial approaches to flat space holography. The former works with corner charges that remain conserved under evolution in Bondi time $u$. The latter works with modes of chiral currents of a 2D celestial CFT. The two types of charges are related by a change of the 2-cycle along which we integrate our conserved 2-form currents. So the formalism of 1-form symmetries provides a clean (non-twistorial) proof of the equivalence of the Carrollian and celestial approaches to large gauge charges.


\section{Discussion}
\label{sec:discussion}

The celestial $S$-algebra governs the asymptotic symmetries of Yang-Mills theory. In this work, we have found that in the self-dual sector of Yang-Mills, every element of the $S$-algebra extends into the bulk to yield a 1-form symmetry.

\begin{figure}[t]
    \centering
    \includegraphics[width=0.65\textwidth]{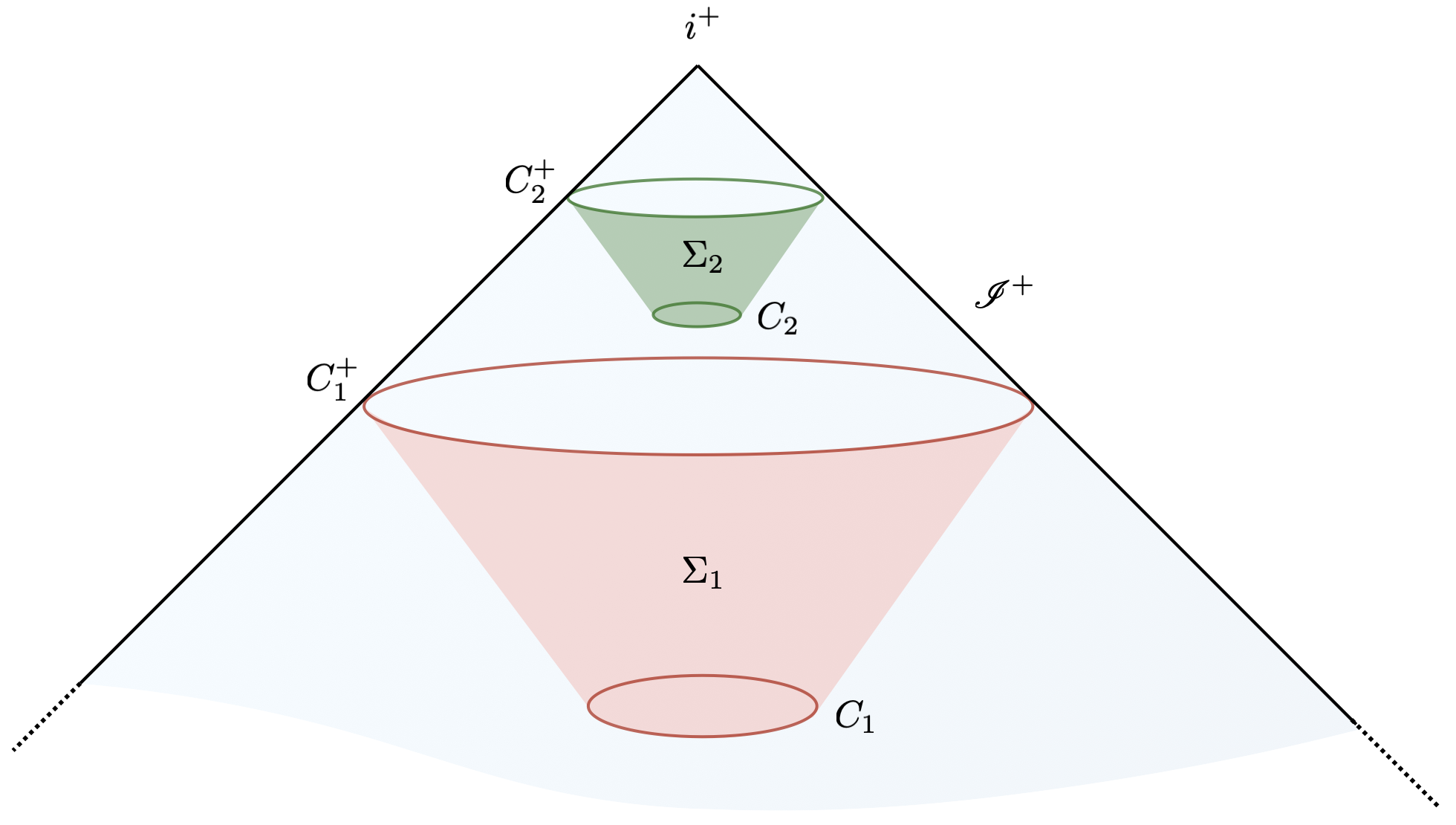}
    \caption{1-form symmetry charges evaluated along the codimension-2 surfaces $C_i$ are anchored to celestial spheres $C_i^+$ through the codimension-1 surfaces $\Sigma_i$. A simple choice of $\Sigma_i$ is obtained as the surface formed from null geodesics emanating from points $(u,v,z,\bz)\in C_i$ along which one holds $(u,z,\bz)$ fixed and lets $v$ vary. Noncommutativity of the asymptotic symmetry charges then directly translates into noncommutativity of our 1-form symmetry generators.}
    \label{noncommute}
\end{figure}

To understand the noncommutativity of our 1-form symmetry charges, one has to recall that the conservation law $\d \bJ_\tau=0$ is valid only after imposition of the sd constraints that generate the gauge symmetry. 
In order to compute the bracket of two charges associated with two different surfaces $C_i$, we need to express each charge as the integral over a codimension-$1$ manifold   $\Sigma_i$ with boundary $C_i \subset \partial \Sigma_i$ \cite{Cattaneo:2016zsq}. In order to construct a nonzero charge, we choose $\Sigma_i$ to be the future null cone with section $C_i$ as in figure \ref{noncommute}. This cone intersects future null infinity $\scri^+$ along a light cone cut $C_i^+$. The conservation law implies that $Q_{\tau_i}(C_i) = Q_{\tau_i}(C_i^+)$, and  the charge  $Q_{\tau_i}(C_i^+)$  can now be interpreted as a 0-form symmetry charge from the asymptotic boundary perspective. The fact that these charges do not commute follows now from the Noether analysis along null infinity developed in \cite{ Cresto:2025bfo, Cresto:2024mne}. 
In geometrical terms, the anchoring of the charge at infinity implies that in order to commute the charges associated with $C_1$ and $C_2$, the celestial cuts $C_i^+$ on $\scri^+$ need to cross.

The key element of this analysis relies on the fact that we can express the charge $Q_\tau(C^+)$ as the integral along an interval $I$ on $\scri^+$, where $\partial I = C^+\cup i^+$, where $i^+$ denotes timelike infinity: 
$ Q_\tau(C^+)=\int_{\mathcal{\scri^+}}
\sum_{s\in\Z}\Tr\big[\big(\Dbar\tau^I_s - \Dhat\tau^I_{s-1}\big)\wedge B_s\big] $
where $\tau^I_s=\chi_I \tau_s$, with $\chi_I$ the characteristic function of the interval $I$ and $\tau_s$ a solution of $\Dbar\tau_s - \Dhat\tau_{s-1}=0$. The noncommutativity of these charges follows from the noncommutativity of  $ (B,A)$  along  $\scri^+$. 
The bracket of two such observables can be evaluated easily when the labels $\tau $  are free (\emph{i.e.}, not field dependent),
\be 
\{ Q_\tau(C^+_1), Q_{\tau'}(C^+_2)\} =
Q_{[\tau,\tau']}(C^+_1)\,,
\ee 
where we assume that $I_1\subset I_2$. A similar commutator of the charges is obtained for gauge parameters $\tau$ which are field dependent, where the Lie bracket is simply replaced by an algebroid bracket $[\cdot, \cdot]\to [\![\cdot,\cdot]\!]$ \cite{Barnich:2010eb, Cresto:2025bfo, Cresto:2024mne}. To conclude, one establishes that the algebroid bracket preserves the dual conditions \eqref{dualrec} and therefore the Poisson bracket realizes the S-algebra. 

The noncommutativity of the charges is therefore due to the fact that they are nonlocal objects that are anchored at infinity and depend implicitly on $\Sigma_i$. It would be desirable to make some of these ideas more explicit.

It would also be highly interesting to see if some of our ideas survive in self-dual gravity (sdGR). One usually does not expect to find any global symmetries in theories of quantum gravity. Nonetheless, the integrability and twistor description of sdGR again lead to a recursion operator and hierarchies of commuting flows on its moduli space of solutions \cite{Dunajski:2000iq}. This often requires working with gauge-fixed descriptions of sdGR in terms of the heavenly equations \cite{Plebanski:1975wn}. It is likely that these gauge-fixed forms of sdGR will again display an infinite class of 2-form symmetries that asymptotically reduce to the $Lw_{1+\infty}$ algebra governing graviton scattering amplitudes \cite{Guevara:2021abz,Strominger:2021mtt,Adamo:2021lrv}. The Noether charges of the $Lw_{1+\infty}$ algebra have also been studied in great detail in \cite{Freidel:2021dfs,Freidel:2021ytz,Donnay:2024qwq,Kmec:2024nmu,Cresto:2024fhd,Cresto:2024mne,Cresto:2025ubl}, and should provide a useful starting point for future investigations.


\acknowledgments

We are grateful to Davide Gaiotto and Max H\"ubner for helpful conversations. We thank Lionel Mason and Romain Ruzziconi for discussions about upcoming parallel work \cite{Kmec:future} in the context of self-dual gravity. A.S.\ is supported by the Gordon and Betty Moore Foundation and the John Templeton Foundation via the Black Hole Initiative.  Research at Perimeter Institute
is supported in part by the Government of Canada through the Department of Innovation, Science
and Economic Development and by the Province of Ontario through the Ministry of Colleges and
Universities. This work is supported by the Simons Collaboration on Celestial
Holography.

\bibliographystyle{JHEP}
\bibliography{corner}

@article{Gaiotto:2017yup,
    author = "Gaiotto, Davide and Kapustin, Anton and Komargodski, Zohar and Seiberg, Nathan",
    title = "{Theta, Time Reversal, and Temperature}",
    eprint = "1703.00501",
    archivePrefix = "arXiv",
    primaryClass = "hep-th",
    doi = "10.1007/JHEP05(2017)091",
    journal = "JHEP",
    volume = "05",
    pages = "091",
    year = "2017"
}

@article{Komargodski:2020mxz,
    author = "Komargodski, Zohar and Ohmori, Kantaro and Roumpedakis, Konstantinos and Seifnashri, Sahand",
    title = "{Symmetries and strings of adjoint QCD$_{2}$}",
    eprint = "2008.07567",
    archivePrefix = "arXiv",
    primaryClass = "hep-th",
    reportNumber = "YITP-SB-20-28",
    doi = "10.1007/JHEP03(2021)103",
    journal = "JHEP",
    volume = "03",
    pages = "103",
    year = "2021"
}

@book{Strominger:2017zoo,
    author = "Strominger, Andrew",
    title = "{Lectures on the Infrared Structure of Gravity and Gauge Theory}",
    eprint = "1703.05448",
    archivePrefix = "arXiv",
    primaryClass = "hep-th",
    isbn = "978-0-691-17973-5",
    publisher = "Princeton University Press",
    year = "2018"
}

@article{Brown:1986nw,
    author = "Brown, J. David and Henneaux, M.",
    title = "{Central Charges in the Canonical Realization of Asymptotic Symmetries: An Example from Three-Dimensional Gravity}",
    doi = "10.1007/BF01211590",
    journal = "Commun. Math. Phys.",
    volume = "104",
    pages = "207--226",
    year = "1986"
}

@article{Strominger:1997eq,
    author = "Strominger, Andrew",
    title = "{Black hole entropy from near horizon microstates}",
    eprint = "hep-th/9712251",
    archivePrefix = "arXiv",
    reportNumber = "HUTP-97-A106",
    doi = "10.1088/1126-6708/1998/02/009",
    journal = "JHEP",
    volume = "02",
    pages = "009",
    year = "1998"
}

@book{Mason:1991rf,
    author = "Mason, L. J. and Woodhouse, N. M. J.",
    title = "{Integrability, selfduality, and twistor theory}",
    year = "1991"
}

@article{Penrose:1972ia,
    author = "Penrose, R. and MacCallum, Malcolm A. H.",
    title = "{Twistor theory: An Approach to the quantization of fields and space-time}",
    doi = "10.1016/0370-1573(73)90008-2",
    journal = "Phys. Rept.",
    volume = "6",
    pages = "241--316",
    year = "1972"
}

@article{He:2014cra,
    author = "He, Temple and Mitra, Prahar and Porfyriadis, Achilleas P. and Strominger, Andrew",
    title = "{New Symmetries of Massless QED}",
    eprint = "1407.3789",
    archivePrefix = "arXiv",
    primaryClass = "hep-th",
    doi = "10.1007/JHEP10(2014)112",
    journal = "JHEP",
    volume = "10",
    pages = "112",
    year = "2014"
}

@article{Eastwood:1981jy,
    author = "Eastwood, Michael G. and Penrose, R. and Wells, R. O.",
    title = "{Cohomology and Massless Fields}",
    doi = "10.1007/BF01942327",
    journal = "Commun. Math. Phys.",
    volume = "78",
    pages = "305--351",
    year = "1981"
}

@book{Ward:1990vs,
    author = "Ward, R. S. and Wells, R. O.",
    title = "{Twistor geometry and field theory}",
    doi = "10.1017/CBO9780511524493",
    isbn = "978-0-521-42268-0, 978-0-521-42268-0, 978-0-511-86977-8",
    publisher = "Cambridge University Press",
    series = "Cambridge Monographs on Mathematical Physics",
    month = "8",
    year = "1991"
}

@article{Kmec:2025ftx,
    author = "Kmec, Adam and Mason, Lionel and Ruzziconi, Romain and Sharma, Atul",
    title = "{S-algebra in gauge theory: twistor, spacetime and holographic perspectives}",
    eprint = "2506.01888",
    archivePrefix = "arXiv",
    primaryClass = "hep-th",
    doi = "10.1088/1361-6382/ae0673",
    journal = "Class. Quant. Grav.",
    volume = "42",
    number = "19",
    pages = "195008",
    year = "2025"
}

@misc{verbitsky2006,
      title="{Quaternionic Dolbeault complex and vanishing theorems on hyperkahler manifolds}", 
      author={Misha Verbitsky},
      year={2006},
      eprint={math/0604303},
      archivePrefix={arXiv},
      primaryClass={math.AG},
      url={https://arxiv.org/abs/math/0604303}, 
}

@article{Dunajski:2000iq,
    author = "Dunajski, Maciej and Mason, L. J.",
    title = "{HyperKahler hierarchies and their twistor theory}",
    eprint = "math/0001008",
    archivePrefix = "arXiv",
    doi = "10.1007/PL00005532",
    journal = "Commun. Math. Phys.",
    volume = "213",
    pages = "641--672",
    year = "2000"
}

@article{Freidel:2023gue,
    author = "Freidel, Laurent and Pranzetti, Daniele and Raclariu, Ana-Maria",
    title = "{On infinite symmetry algebras in Yang-Mills theory}",
    eprint = "2306.02373",
    archivePrefix = "arXiv",
    primaryClass = "hep-th",
    doi = "10.1007/JHEP12(2023)009",
    journal = "JHEP",
    volume = "12",
    pages = "009",
    year = "2023"
}

@article{Strominger:2021mtt,
    author = "Strominger, Andrew",
    title = "{$w_{1+\infty}$ Algebra and the Celestial Sphere: Infinite Towers of Soft Graviton, Photon, and Gluon Symmetries}",
    eprint = "2105.14346",
    archivePrefix = "arXiv",
    primaryClass = "hep-th",
    doi = "10.1103/PhysRevLett.127.221601",
    journal = "Phys. Rev. Lett.",
    volume = "127",
    number = "22",
    pages = "221601",
    year = "2021"
}

@article{Guevara:2021abz,
    author = "Guevara, Alfredo and Himwich, Elizabeth and Pate, Monica and Strominger, Andrew",
    title = "{Holographic symmetry algebras for gauge theory and gravity}",
    eprint = "2103.03961",
    archivePrefix = "arXiv",
    primaryClass = "hep-th",
    doi = "10.1007/JHEP11(2021)152",
    journal = "JHEP",
    volume = "11",
    pages = "152",
    year = "2021"
}

@article{Cresto:2025bfo,
    author = "Cresto, Nicolas",
    title = "{Asymptotic higher spin symmetries III: Noether realization in Yang{\textendash}Mills theory}",
    eprint = "2501.08856",
    archivePrefix = "arXiv",
    primaryClass = "hep-th",
    doi = "10.1007/s11005-025-02027-7",
    journal = "Lett. Math. Phys.",
    volume = "115",
    number = "6",
    pages = "133",
    year = "2025"
}

@article{Pasterski:2022jzc,
    author = "Pasterski, Sabrina",
    title = "{A shorter path to celestial currents}",
    eprint = "2201.06805",
    archivePrefix = "arXiv",
    primaryClass = "hep-th",
    doi = "10.1007/JHEP05(2023)190",
    journal = "JHEP",
    volume = "05",
    pages = "190",
    year = "2023"
}

@article{Crawley:2021ivb,
    author = "Crawley, Erin and Miller, Noah and Narayanan, Sruthi A. and Strominger, Andrew",
    title = "{State-operator correspondence in celestial conformal field theory}",
    eprint = "2105.00331",
    archivePrefix = "arXiv",
    primaryClass = "hep-th",
    doi = "10.1007/JHEP09(2021)132",
    journal = "JHEP",
    volume = "09",
    pages = "132",
    year = "2021"
}

@inproceedings{Iqbal:2024pee,
    author = "Iqbal, Nabil",
    title = "{Jena lectures on generalized global symmetries: principles and applications}",
    eprint = "2407.20815",
    archivePrefix = "arXiv",
    primaryClass = "hep-th",
    month = "7",
    year = "2024"
}

@article{Gaiotto:2014kfa,
    author = "Gaiotto, Davide and Kapustin, Anton and Seiberg, Nathan and Willett, Brian",
    title = "{Generalized Global Symmetries}",
    eprint = "1412.5148",
    archivePrefix = "arXiv",
    primaryClass = "hep-th",
    doi = "10.1007/JHEP02(2015)172",
    journal = "JHEP",
    volume = "02",
    pages = "172",
    year = "2015"
}

@article{Bhardwaj:2023kri,
    author = "Bhardwaj, Lakshya and Bottini, Lea E. and Fraser-Taliente, Ludovic and Gladden, Liam and Gould, Dewi S. W. and Platschorre, Arthur and Tillim, Hannah",
    title = "{Lectures on generalized symmetries}",
    eprint = "2307.07547",
    archivePrefix = "arXiv",
    primaryClass = "hep-th",
    doi = "10.1016/j.physrep.2023.11.002",
    journal = "Phys. Rept.",
    volume = "1051",
    pages = "1--87",
    year = "2024"
}

@article{Gaiotto:2017zba,
    author = "Gaiotto, Davide and Johnson-Freyd, Theo",
    title = "{Symmetry Protected Topological phases and Generalized Cohomology}",
    eprint = "1712.07950",
    archivePrefix = "arXiv",
    primaryClass = "hep-th",
    doi = "10.1007/JHEP05(2019)007",
    journal = "JHEP",
    volume = "05",
    pages = "007",
    year = "2019"
}

@article{Berean-Dutcher:2025ohp,
    author = "Berean-Dutcher, Jonah and Derda, Maria and Parra-Martinez, Julio",
    title = "{Soft theorems from higher symmetries}",
    eprint = "2505.03566",
    archivePrefix = "arXiv",
    primaryClass = "hep-th",
    reportNumber = "CALT-TH 2025-013",
    doi = "10.1007/JHEP03(2026)193",
    journal = "JHEP",
    volume = "03",
    pages = "193",
    year = "2026"
}

@article{Tizzano:2026rgr,
    author = "Tizzano, Luigi",
    title = "{Comments on Symmetry Operators, Asymptotic Charges and Soft Theorems}",
    eprint = "2604.06088",
    archivePrefix = "arXiv",
    primaryClass = "hep-th",
    reportNumber = "CERN-TH-2026-083",
    month = "4",
    year = "2026"
}

@article{Barnich:2010eb,
    author = "Barnich, Glenn and Troessaert, Cedric",
    title = "{Aspects of the BMS/CFT correspondence}",
    eprint = "1001.1541",
    archivePrefix = "arXiv",
    primaryClass = "hep-th",
    reportNumber = "ULB-TH-09-28",
    doi = "10.1007/JHEP05(2010)062",
    journal = "JHEP",
    volume = "05",
    pages = "062",
    year = "2010"
}

@article{Cresto:2024mne,
    author = "Cresto, Nicolas and Freidel, Laurent",
    title = "{Asymptotic higher spin symmetries II: Noether realization in gravity}",
    eprint = "2410.15219",
    archivePrefix = "arXiv",
    primaryClass = "hep-th",
    doi = "10.1007/JHEP03(2026)147",
    journal = "JHEP",
    volume = "03",
    pages = "147",
    year = "2026"
}

@article{Cattaneo:2016zsq,
    author = "Cattaneo, Alberto S. and Perez, Alejandro",
    title = "{A note on the Poisson bracket of 2d smeared fluxes in loop quantum gravity}",
    eprint = "1611.08394",
    archivePrefix = "arXiv",
    primaryClass = "gr-qc",
    doi = "10.1088/1361-6382/aa69b4",
    journal = "Class. Quant. Grav.",
    volume = "34",
    number = "10",
    pages = "107001",
    year = "2017"
}

@article{Donnelly:2016auv,
    author = "Donnelly, William and Freidel, Laurent",
    title = "{Local subsystems in gauge theory and gravity}",
    eprint = "1601.04744",
    archivePrefix = "arXiv",
    primaryClass = "hep-th",
    doi = "10.1007/JHEP09(2016)102",
    journal = "JHEP",
    volume = "09",
    pages = "102",
    year = "2016"
}

@article{Ciambelli:2022vot,
    author = "Ciambelli, Luca",
    title = "{From Asymptotic Symmetries to the Corner Proposal}",
    eprint = "2212.13644",
    archivePrefix = "arXiv",
    primaryClass = "hep-th",
    doi = "10.22323/1.435.0002",
    journal = "PoS",
    volume = "Modave2022",
    pages = "002",
    year = "2023"
}

@inproceedings{Pasterski:2021raf,
    author = "Pasterski, Sabrina and Pate, Monica and Raclariu, Ana-Maria",
    title = "{Celestial Holography}",
    booktitle = "{Snowmass 2021}",
    eprint = "2111.11392",
    archivePrefix = "arXiv",
    primaryClass = "hep-th",
    month = "11",
    year = "2021"
}

@article{Ruzziconi:2026bix,
    author = "Ruzziconi, Romain",
    title = "{Carrollian Physics and Holography}",
    eprint = "2602.02644",
    archivePrefix = "arXiv",
    primaryClass = "hep-th",
    month = "2",
    year = "2026"
}

@article{Pano:2023slc,
    author = "Pano, Yorgo and Puhm, Andrea and Trevisani, Emilio",
    title = "{Symmetries in Celestial CFT$_{d}$}",
    eprint = "2302.10222",
    archivePrefix = "arXiv",
    primaryClass = "hep-th",
    reportNumber = "CPHT-RR071.122022",
    doi = "10.1007/JHEP07(2023)076",
    journal = "JHEP",
    volume = "07",
    pages = "076",
    year = "2023"
}

@article{Donnay:2022aba,
    author = "Donnay, Laura and Fiorucci, Adrien and Herfray, Yannick and Ruzziconi, Romain",
    title = "{Carrollian Perspective on Celestial Holography}",
    eprint = "2202.04702",
    archivePrefix = "arXiv",
    primaryClass = "hep-th",
    doi = "10.1103/PhysRevLett.129.071602",
    journal = "Phys. Rev. Lett.",
    volume = "129",
    number = "7",
    pages = "071602",
    year = "2022"
}

@article{Donnay:2022wvx,
    author = "Donnay, Laura and Fiorucci, Adrien and Herfray, Yannick and Ruzziconi, Romain",
    title = "{Bridging Carrollian and celestial holography}",
    eprint = "2212.12553",
    archivePrefix = "arXiv",
    primaryClass = "hep-th",
    doi = "10.1103/PhysRevD.107.126027",
    journal = "Phys. Rev. D",
    volume = "107",
    number = "12",
    pages = "126027",
    year = "2023"
}

@article{Freidel:2021ytz,
    author = "Freidel, Laurent and Pranzetti, Daniele and Raclariu, Ana-Maria",
    title = "{Higher spin dynamics in gravity and w1+{\ensuremath{\infty}} celestial symmetries}",
    eprint = "2112.15573",
    archivePrefix = "arXiv",
    primaryClass = "hep-th",
    doi = "10.1103/PhysRevD.106.086013",
    journal = "Phys. Rev. D",
    volume = "106",
    number = "8",
    pages = "086013",
    year = "2022"
}

@article{Cresto:2025ubl,
    author = "Cresto, Nicolas and Freidel, Laurent",
    title = "{Asymptotic higher spin symmetries. Part IV. Einstein-Yang-Mills theory}",
    eprint = "2505.04327",
    archivePrefix = "arXiv",
    primaryClass = "hep-th",
    doi = "10.1007/JHEP12(2025)097",
    journal = "JHEP",
    volume = "12",
    pages = "097",
    year = "2025"
}

@article{Strominger:2013jfa,
    author = "Strominger, Andrew",
    title = "{On BMS Invariance of Gravitational Scattering}",
    eprint = "1312.2229",
    archivePrefix = "arXiv",
    primaryClass = "hep-th",
    doi = "10.1007/JHEP07(2014)152",
    journal = "JHEP",
    volume = "07",
    pages = "152",
    year = "2014"
}

@article{Plebanski:1975wn,
    author = "Plebanski, J. F.",
    title = "{Some solutions of complex Einstein equations}",
    doi = "10.1063/1.522505",
    journal = "J. Math. Phys.",
    volume = "16",
    pages = "2395--2402",
    year = "1975"
}

@article{Adamo:2021lrv,
    author = "Adamo, Tim and Mason, Lionel and Sharma, Atul",
    title = "{Celestial $w_{1+\infty}$ Symmetries from Twistor Space}",
    eprint = "2110.06066",
    archivePrefix = "arXiv",
    primaryClass = "hep-th",
    doi = "10.3842/SIGMA.2022.016",
    journal = "SIGMA",
    volume = "18",
    pages = "016",
    year = "2022"
}

@article{Freidel:2021dfs,
    author = "Freidel, Laurent and Pranzetti, Daniele and Raclariu, Ana-Maria",
    title = "{Sub-subleading soft graviton theorem from asymptotic Einstein{\textquoteright}s equations}",
    eprint = "2111.15607",
    archivePrefix = "arXiv",
    primaryClass = "hep-th",
    doi = "10.1007/JHEP05(2022)186",
    journal = "JHEP",
    volume = "05",
    pages = "186",
    year = "2022"
}

@article{Donnay:2024qwq,
    author = "Donnay, Laura and Freidel, Laurent and Herfray, Yannick",
    title = "{Carrollian $\mathscr Lw_{1+\infty}$ representation from twistor space}",
    eprint = "2402.00688",
    archivePrefix = "arXiv",
    primaryClass = "hep-th",
    doi = "10.21468/SciPostPhys.17.4.118",
    journal = "SciPost Phys.",
    volume = "17",
    number = "4",
    pages = "118",
    year = "2024"
}

@article{Cresto:2024fhd,
    author = "Cresto, Nicolas and Freidel, Laurent",
    title = "{Asymptotic higher spin symmetries I: covariant wedge algebra in gravity}",
    eprint = "2409.12178",
    archivePrefix = "arXiv",
    primaryClass = "hep-th",
    doi = "10.1007/s11005-025-01921-4",
    journal = "Lett. Math. Phys.",
    volume = "115",
    number = "2",
    pages = "39",
    year = "2025"
}

@article{Kmec:2024nmu,
    author = "Kmec, Adam and Mason, Lionel and Ruzziconi, Romain and Yelleshpur Srikant, Akshay",
    title = "{Celestial $Lw_{1+\infty}$ charges from a twistor action}",
    eprint = "2407.04028",
    archivePrefix = "arXiv",
    primaryClass = "hep-th",
    doi = "10.1007/JHEP10(2024)250",
    journal = "JHEP",
    volume = "10",
    pages = "250",
    year = "2024"
}

@article{Kmec:future,
    author = "Kmec, Adam and Mason, Lionel and Ruzziconi, Romain",
    title = "{Quasi-local celestial charges and multipoles, to appear}",
    year = "to appear"
}

\end{document}